# Using Lotkaian Informetrics for Ranking in Digital Libraries

Philipp Schaer

GESIS – Leibniz Institute for the Social Sciences, Lennéstr. 30, 53113, Germany
philipp.schaer@gesis.org

**Abstract.** The purpose of this paper is to propose the use of models, theories and laws in bibliometrics and scientometrics to enhance information retrieval processes, especially ranking. A common pattern in many man-made data sets is Lotka's Law which follows the well-known power-law distributions. These informetric distributions can be used to give an alternative order to large and scattered result sets and can be applied as a new ranking mechanism. The polyrepresentation of information in Digital Library systems is used to enhance the retrieval quality, to overcome the drawbacks of the typical term-based ranking approaches and to enable users to explore retrieved document sets from a different perspective.

**Keywords:** Lotka's Law, informetrics, scientometrics, information retrieval, non-textual ranking, Digital Libraries

## 1 Introduction

When searching in a typical metadata-driven Digital Library (DL) users face more or less sophisticated search forms that might retrieve a vast amount of documents. The result set is often too huge and heterogeneous to be overlooked completely. One of the most important tasks of an interactive information retrieval (IIR, [1]) system therefore is to provide a rational and useful ranking of the result set to allow the user to find the most relevant documents right at the top of a result list. Traditional term-based ranking methods like *tf-idf* have their weaknesses and the improvements seem to stagnate [2].

Therefore the aim of this paper is to outline an approach to enhance the retrieval quality in DL systems by looking beyond simple term-based ranking. In section 2 the principle of Lotkaian Informetrics and its combination with term-based ranking methods is described. This approach is evaluated in section 3 and the effects of this approach are discussed in section 4.

## 2 Combining Lotkaian Informetrics and Term-based Ranking

A typical DL's metadata structure holds many additional representations of information objects. An entry on a book might contain the title, the authors, or the full text.

Additionally a lot of structural and functional data is included like the authors' affiliations, the publisher, the place the book was printed, descriptors from a controlled vocabulary etc. According to Hjørland the extension of retrieval functions through intellectually or automatically generated metadata is seen as one of the key roles of documentation and information science (LIS) [3]. Its function is not only to build up and maintain document collections but to create added values for the user [4].

One of the central principles in LIS is the so-called power-law distribution. These statistical distributions can be seen in bibliometric analyses of journal publications, the use of language and term frequencies [5]. Many man-made or naturally occurring phenomena show typical features that follow equation 1:

$$f(x) = cx^{-\alpha}. \tag{1}$$

Where c is constant, x corresponds to the rank of the actual entity and $\alpha$ normally is between 1 and 2. Power-laws are monotone, so they rise or fall continuously. When power-laws are used to describe distributions $\alpha$ is most likely to be positive. Generally speaking this says that large events are rare, but small ones are quite common. For example there are only few words, like "and" and "the" that occur very frequently, but many which occur rarely. When these assumptions hold true a typical plot of these distributions look like those on figure 1. We can see the typical long tail. When plotted on a double logarithmic scale all distributions appear as a straight line. In the information production process these observations are called Lotkaian Informetrics [6], following the well-known studies of Lotka.

One well-evaluated type of a power-law distribution is Bradford's Law, which was utilized for ranking in the form of Bradfordizing [7] where documents that belong to a so-called core journal are favored during the ranking process. One problem of this approach is that core journal documents don't have an inner ranking, which distinguishes them from other documents belonging to the same journal. In a typical multi-database scenario potentially hundreds of documents might be ranked higher then those belonging to a peripheral journal and the benefit of the re-ranking is marginalized. Since there is no distinction in the score of the single documents a user has to evaluate each single document on its relevancy.

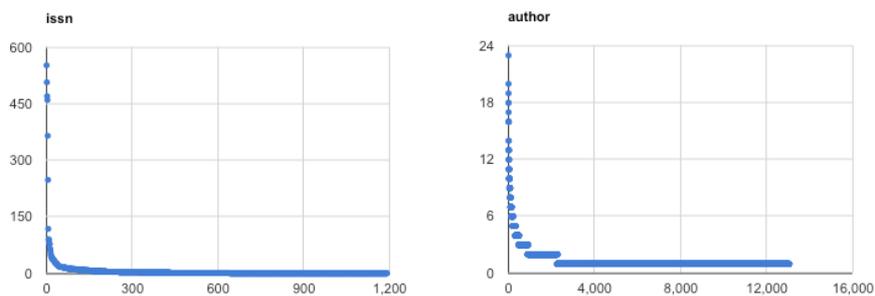

**Fig. 1.** Rank-frequency plot for the result set on an query on "violence AND family" on six domain-specific literature databases (social science, pedagogics and sport science) for journal's ISSN (Bradford's Law) and author names (Lotka's Law).

Therefore a combination of traditional term-based ranking like *tf-idf* and a further component that represents additional ranking-relevant information is suggested. The combination of the term-based scoring and the bibliometric factor can be utilized for the final ranking:

$$\text{Score(q,d)} = \sum_{t \in q} tf_{t,d} \log \frac{N}{df_t} \left( \sum_{q \in N} \frac{ft_d}{N} \right)^k .$$ (2)

The first part represents the *tf-idf* score for a query q and a document d in a total result set N. The second part of the equation represents a sum on the number of documents belonging to the actual entity in consideration – in this example from the same journal. The exponent k can be used as a weighing factor.

## 3 Experiment

The effects of the proposed method shown in formula 2 are evaluated with the GIRT-4 evaluation dataset, which was used in the CLEF campaign. The dataset was enriched with the journal's ISSN. For this experiment we used the CLEF topic files 126 to 150 and tested three different ranking implementations: standard *tf-idf* ranking, re-ranking according to journal frequency (comparable to Bradfordizing) and re-ranking according to author frequency (inspired by Lotka's Law). In figure 2 we see the precision values for the three implementations for the first 5, 10, 20, 30 and 100 documents (p@5 to p@100).

While re-ranking with respect to the author frequency was better than the *tf-idf* baseline for p@5 to p@30 the journal frequencies can only produce better results in for p@5 and p@10. Comparing the top 10 documents from each result set only 3.3 documents (in average) intersect.

The number of documents returned from each retrieval method was very biased: While the baseline set (*tf-idf* for all 25 topics) contained 1077 documents of which 638 were relevant, the author re-rank set only contained 423 relevant out of 561 retrieved and the journal re-rank set contained 239 out of 364 documents respectively.

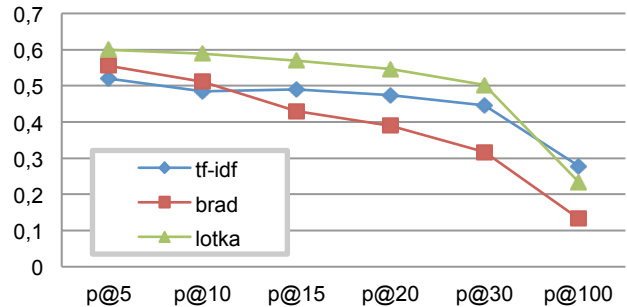

**Fig. 2.** Precision values for three ranking methods: standard *tf-idf* ranking (tf-idf), re-ranking according to journal frequency (brad) and re-ranking according to author frequency (lotka).

## 4 Discussion and Outlook

Commercial web search engines have used non-textual rankings approaches for a long time. Examples are the Google PageRank or the citation count from Google Scholar. The common idea behind these and the shown ranking method is their bibliometric-inspired approach. Non-textual ranking methods in the context of domain specific Digital Libraries [8] were positively evaluated before and two outcomes correspond to these previously observed results: (1) precision values in the first 10 to 20 documents are better for some types of re-ranking (author frequency in this case) and (2) the top documents per re-ranking mechanism are mostly disjoint. One can interpret this as an alternative perspective on the data set.

The precision values have to be seen in respect to the unequal sizes of the different retrieved sets. Since the journal re-rank can only be applied to journal articles all monographs and other literature which is included in the GIRT set is discarded. This is the same to the author re-rank were every document with a missing author name is ignored. This is an open issue and has to be analyzed further.

Since the GIRT dataset was enriched with additional bibliography metadata (publishers, affiliations, locations etc.) the experiments can be repeated with completely different entities. Various re-ranking scenarios can be tested and promising candidates can be identified. These candidates could be used to implement a "mainstream-rank" or an "outsider-rank", according to the user's interests. Experts in a field might be more interested in documents which are located on the long-tail (cf. figure 1), while novice users might prefer to re-rank result sets according to important authors, publishers etc. Further user-centered evaluation is essential here.

## References

1. Fuhr, N.: A probability ranking principle for interactive information retrieval. Inf Retrieval. 11, 251-265 (2008).
2. Armstrong, T.G., Moffat, A., Webber, W., Zobel, J.: Improvements that don't add up: ad-hoc retrieval results since 1998. In: Cheung, D.W.-L., Song, I.-Y., Chu, W.W., Hu, X., und Lin, J.J. (eds.) CIKM. S. 601-610. ACM (2009).
3. Hjørland, B.: Library and information science: practice, theory, and philosophical basis. Information Processing and Management. 36, 501-531 (2000).
4. Mayr, P., Mutschke, P., Petras, V., Schaer, P., Sure, Y.: Applying Science Models for Search. Information und Wissen: global, sozial und frei? - Proceedings des 12. Internationalen Symposiums für Informationswissenschaft (ISI 2011). S. 184--196. Verlag Werner Hülsbusch, Boizenburg (2011).
5. Newman, M.E.J.: Power laws, Pareto distributions and Zipf's law. Contemporary Physics. 46, 323 - 351 (2005).
6. Egghe, L.: Power Laws in the Information Production Process: Lotkaian Informetrics. Elsevier, Oxford (2005).
7. White, H.D.: „Bradfordizing" search output: how it would help online users. Online Review. 5, 47-54 (1981).
8. Schaer, P., Mayr, P., Mutschke, P.: Implications of Inter-Rater Agreement on a Student Information Retrieval Evaluation. In: Atzmüller, M., Benz, D., Hotho, A., und Stumme, G. (eds.) Proceedings of LWA2010, Kassel, Germany (2010).